\documentclass[aps,pre,twocolumn,groupedaddress,superscriptaddress,showpacs]{revtex4-1}

\usepackage{graphicx}
\usepackage{color}
\usepackage{amsmath}
\usepackage{amsfonts}
\usepackage{amssymb}
\usepackage{amsmath}
\usepackage{dcolumn}

\begin{document}
\title{Bohman--Frieze--Wormald model on the lattice, yielding a discontinuous percolation transition}

\author{K. J. Schrenk}
    \email{jschrenk@ethz.ch}
    \affiliation{Computational Physics for Engineering Materials, IfB, ETH Zurich, Schafmattstrasse 6, CH-8093 Zurich, Switzerland}
    
  \author{A. Felder}
    \email{afelder@ethz.ch}
    \affiliation{Computational Physics for Engineering Materials, IfB, ETH Zurich, Schafmattstrasse 6, CH-8093 Zurich, Switzerland}

  \author{S. Deflorin}
    \email{sdeflori@ethz.ch}
    \affiliation{Computational Physics for Engineering Materials, IfB, ETH Zurich, Schafmattstrasse 6, CH-8093 Zurich, Switzerland}
    
  \author{N. A. M. Ara\'ujo}
    \email{nuno@ethz.ch}
    \affiliation{Computational Physics for Engineering Materials, IfB, ETH Zurich, Schafmattstrasse 6, CH-8093 Zurich, Switzerland}
    
  \author{R. M. D'Souza}
    \email{raissa@cse.ucdavis.edu}
    \affiliation{University of California, Davis, California 95616, USA}
    \affiliation{Santa Fe Institute, 1399 Hyde Park Road, Santa Fe, New Mexico 87501, USA}

  \author{H. J. Herrmann}
	\email{hans@ifb.baug.ethz.ch}
	\affiliation{Computational Physics for Engineering Materials, IfB, ETH Zurich, Schafmattstrasse 6, CH-8093 Zurich, Switzerland}
	\affiliation{Departamento de F\'isica, Universidade Federal do Cear\'a, Campus do Pici, 60451-970 Fortaleza, Cear\'a, Brazil}

\begin{abstract}
The BFW model introduced by Bohman, Frieze, and Wormald [Random Struct. Algorithms, {\bf 25}, 432 (2004)] and recently investigated in the framework of discontinuous percolation by Chen and D'Souza [Phys. Rev. Lett., {\bf 106}, 115701 (2011)], is studied on the square and simple-cubic lattices. In two and three dimensions, we find numerical evidence for a strongly discontinuous transition. In two dimensions, the clusters at the threshold are compact with a fractal surface of fractal dimension $d_f=1.49\pm0.02$. On the simple-cubic lattice, distinct jumps in the size of the largest cluster are observed. We proceed to analyze the tree-like version of the model, where only merging bonds are sampled, for dimension two to seven. The transition is again discontinuous in any considered dimension. Finally, the dependence of the cluster-size distribution at the threshold on the spatial dimension is also investigated.
\end{abstract}

\maketitle

\section{\label{sec::intro}Introduction}
Percolation is a classical model in Statistical Physics which, despite consisting of simple geometrical rules, has found application in a broad range of problems \cite{Stauffer94,Sahimi94}. In classical (random) percolation, a fraction $p$ of sites or bonds are occupied uniformly at random. At a critical occupation fraction $p_c$ the system undergoes a continuous transition from a non-percolating to a percolating state. A recent work by Achlioptas, D'Souza, and Spencer \cite{Achlioptas09} hinted at the possibility of a discontinuous percolation transition by slightly modifying the bond selection rules of evolving graphs, coining the term {\it explosive percolation}.
This surprising result led to intensive research efforts studying the model on different topologies and dimensions, yielding ambiguous results regarding the nature of the transition \cite{Ziff09,Ziff10,Radicchi09,Radicchi10,Fortunato11,Friedman09,Hooyberghs11,Cho09,Cho10,Cho10b,Cho11,Rozenfeld10,Pan10,Bastas11,Choi11}.
More recently, the transition in the original model was formally demonstrated to be continuous in the mean-field limit \cite{Riordan11} and there is evidence that the same applies to other models \cite{daCosta10,Grassberger11,Lee11,Tian10}.
Alternative models have been devised and analyzed in detail \cite{Araujo10,Manna10,Manna11,DSouza10,Christensen11,Andrade11b,Moreira10,Araujo11,Schrenk11,Boettcher11}, several of them with clear signs supporting the hypothesis of a discontinuous transition.

In particular, Chen and D'Souza considered the percolation properties on the random graph of the BFW model \cite{Chen11,Chen11b}, originally introduced by Bohman, Frieze, and Wormald \cite{Bohman04}, reporting a discontinuous transition with several giant components in the thermodynamic limit. In general, the behavior of a system undergoing a transition depends on the dimension \cite{Stanley71}. In this paper, we analyze the original BFW model on the $2D$ square lattice and on the $3D$ cubic lattice, as well as the tree-like version up to dimension seven.

The manuscript is organized as follows. In Sec.\,\ref{sec::2d} we discuss the BFW model, adapted to the square lattice, along with some definitions. Section \ref{sec::3d} contains the extension to three dimensions. The tree-like version is analyzed in Sec.\,\ref{sec::llbfw}. We conclude with the final remarks in Sec.\,\ref{sec::end}.

\section{\label{sec::2d}BFW model on the square lattice}
The BFW model has recently been analyzed by Chen and D'Souza \cite{Chen11,Chen11b} on the random graph, by extending the algorithm initially introduced by Bohman, Frieze, and Wormald \cite{Bohman04}. Here, we consider this model on a square lattice of linear size $L$ with periodic boundary conditions and number of sites $N$, where $N=L^2$. Initially, all sites are occupied and all bonds are unoccupied (empty), such that there are $N$ clusters of unitary size. As in Ref.\,\cite{Chen11,Chen11b}, one occupies the bonds according to the following procedure. Let $u$ be the total number of selected bonds, $t$ the number of occupied bonds, and $k$ the stage of the process, initially set to $k=2$. The first bond is occupied at random, such that $t=u=1$. Then, at each step $u$,
\begin{enumerate}
\item One bond, among the unoccupied ones, is selected uniformly at random.
\item $l$ is measured as the maximum cluster size if the selected bond is occupied.
\item If $l\leq k$, go to $(4)$.\\
	Else if $t/u\geq g(k)=1/2+\sqrt{1/(2k)}$, go to $(5)$.\\
	Else increment $k$ by one and go to $(3)$.
\item Occupy the selected bond.
\item Increment $u$ by one.
\end{enumerate}
\begin{figure}
	\includegraphics[width=\columnwidth]{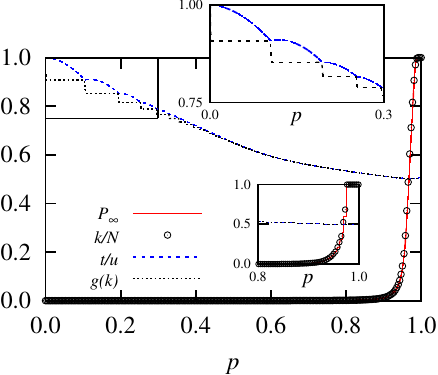}
	\caption{(Color online)
	Fraction of sites in the largest cluster $P_\infty$ (solid red line), stage per site $k/N$ $(\circ)$, occupied bonds per step $t/u$ (dashed upper blue line), and $g(k)$ (dotted lower black line) as a function of the fraction of occupied bonds $p$, for the BFW model on the square lattice. As reported in Ref.~\cite{Bohman04,Chen11} for the random graph, $k$ is essentially equal to $s_\text{max}$. Therefore the corresponding curves in the figure, $P_\infty$ and $k/N$, almost overlap. The upper inset shows a close-up view of the upper-left corner of the main plot, as indicated by the box. The system size is $N=512^2$, the measured values are averages over more than $10^5$ samples. In the lower inset we see the same variables as the main plot, measured for a single realization.
	\label{fig::evo2}
	}
\end{figure}
%
\begin{figure}
	\includegraphics[width=\columnwidth]{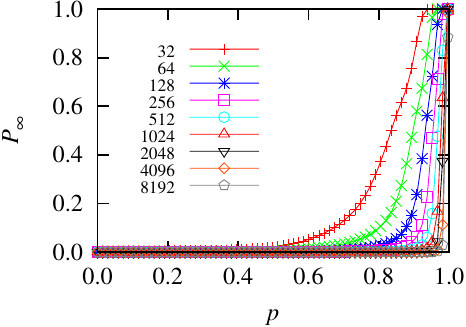}
	\caption{
	(Color online)
	Fraction of sites in the largest cluster $P_\infty$ as a function of the fraction of occupied bonds $p$ on the square lattice for different lattice sizes $L$ (from left to right: $32$, $64$, $128$, $256$, $512$, $1024$, $2048$, $4096$, and $8192$).
	\label{fig::OrderParameterPlot2D}
	}
\end{figure}
%
\begin{figure}
	\begin{tabular}{c}	
		\includegraphics[width=0.6\columnwidth]{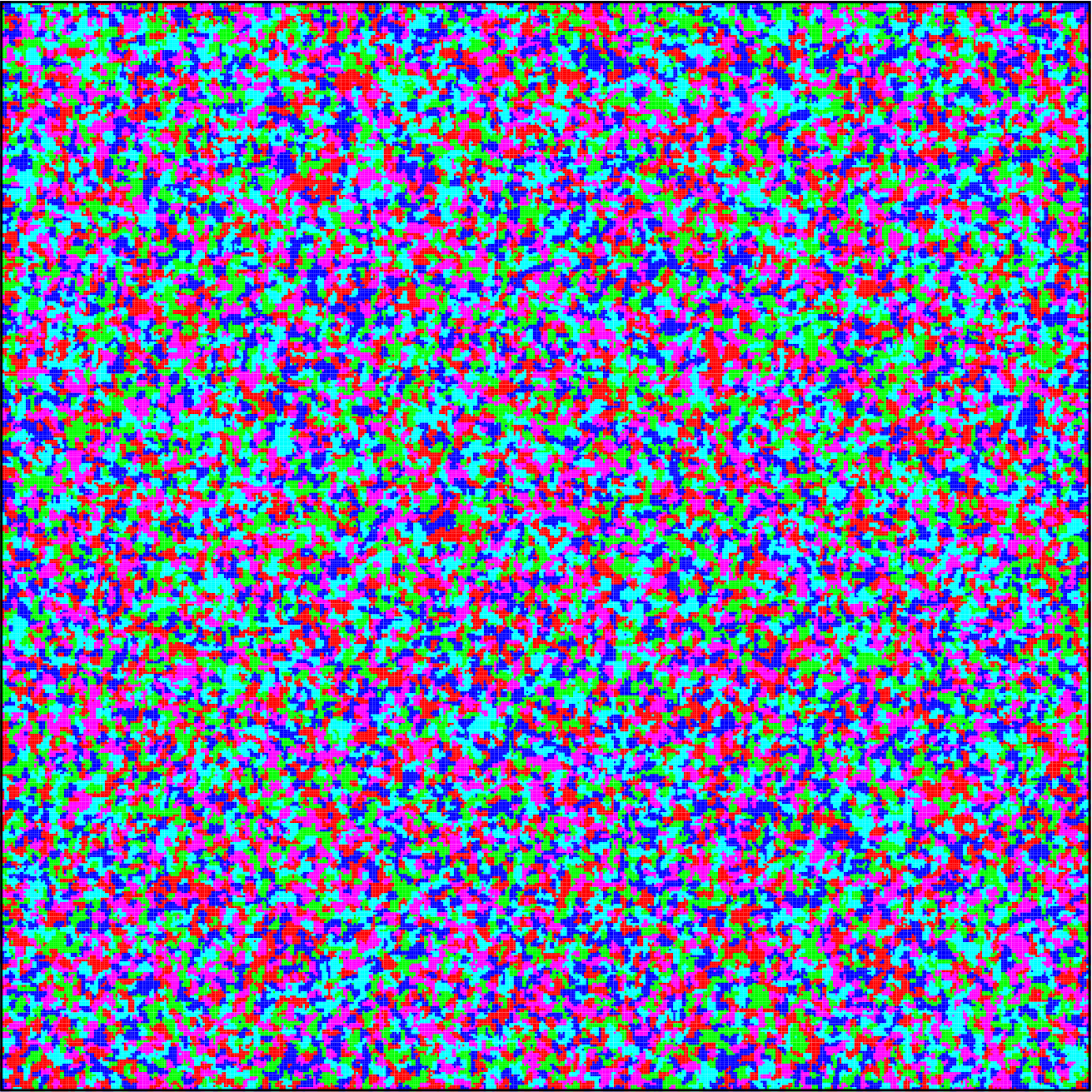}\\
		(a)\\
		\includegraphics[width=0.6\columnwidth]{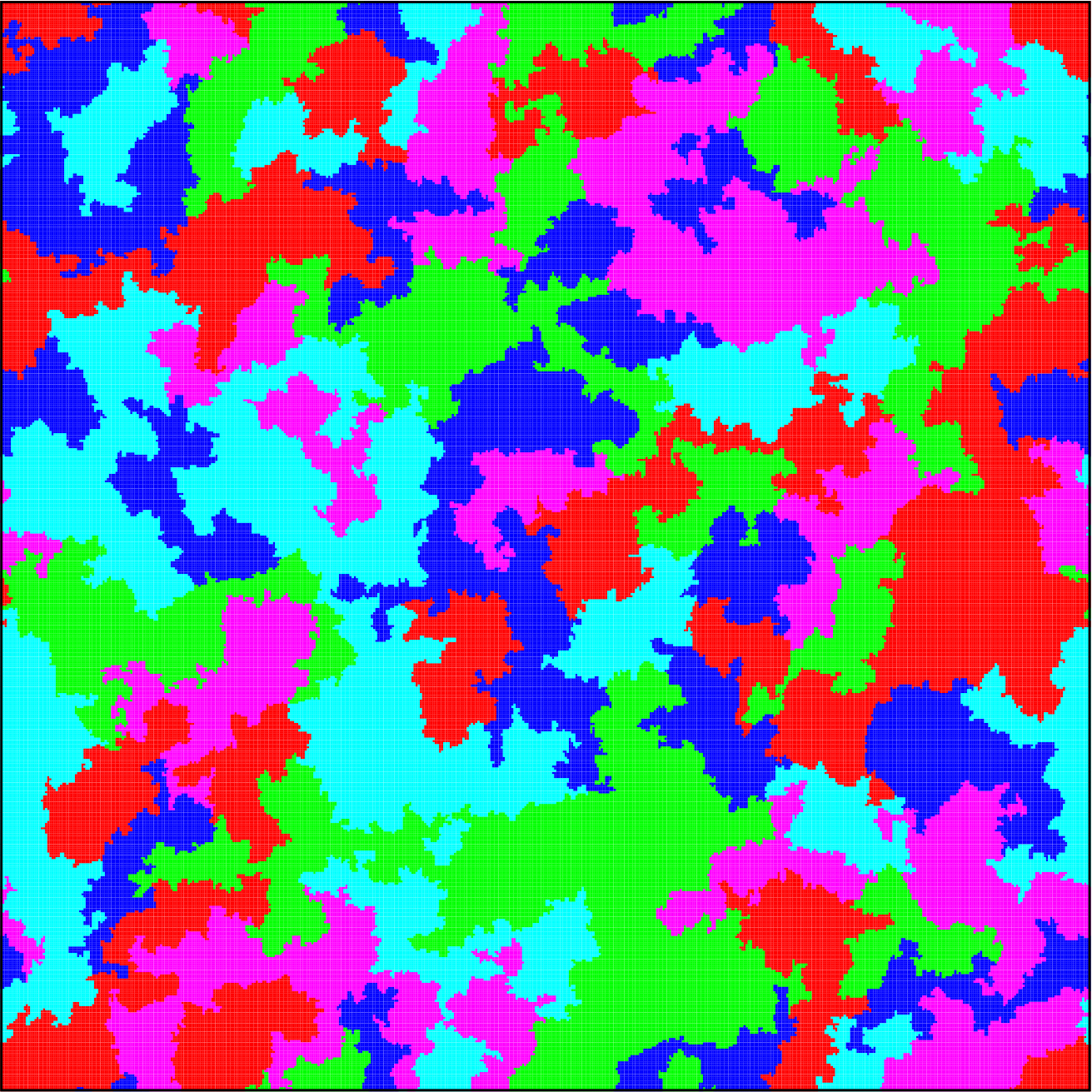}\\
		(b)\\
		\includegraphics[width=0.6\columnwidth]{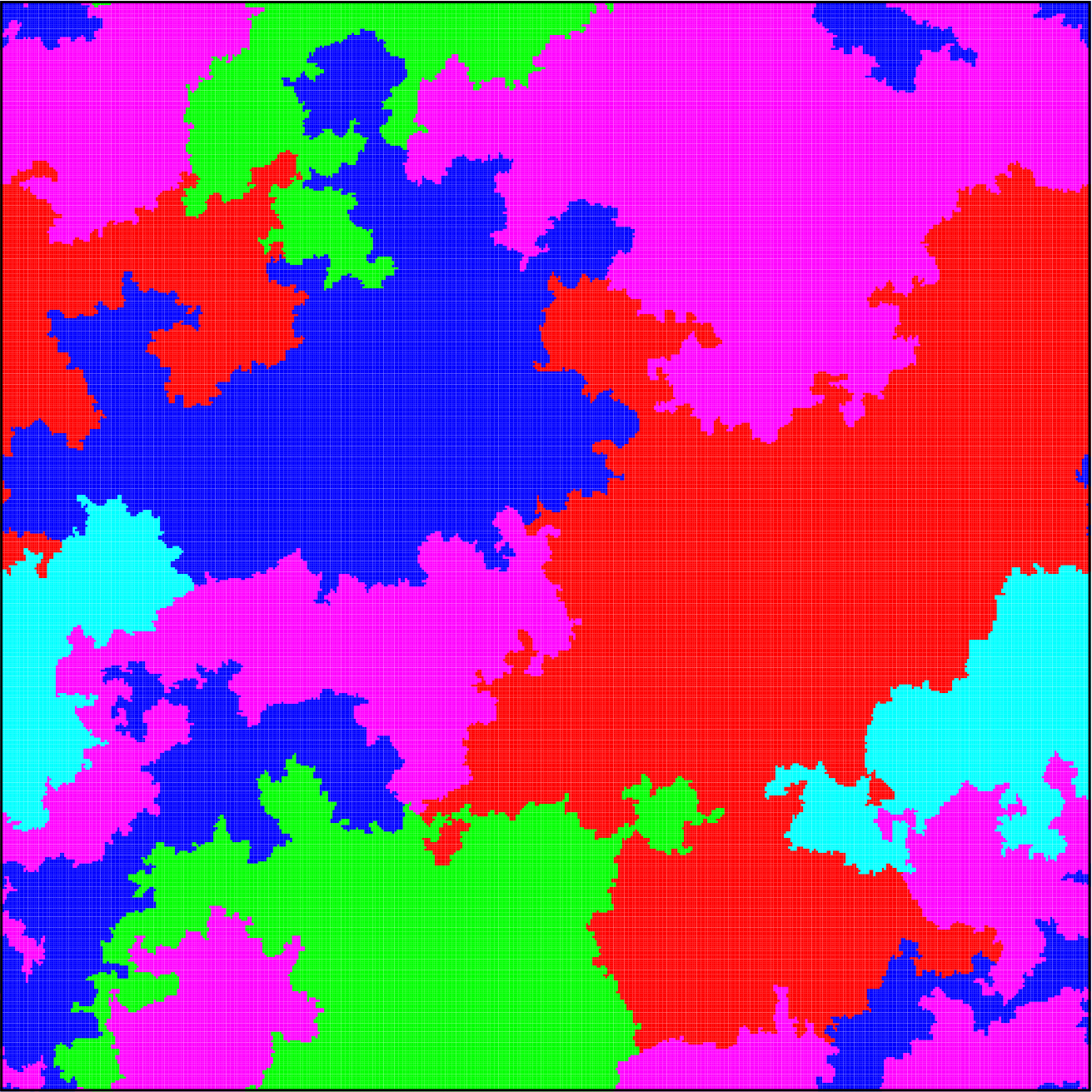}\\
		(c)
	\end{tabular}
		\caption{(Color online)
		Snapshots of the system evolving according to the BFW model, obtained on a square lattice with $512^2$ sites. The following fractions of occupied bonds are shown: (a) $p=0.5$, (b) $0.9$, and (c) $0.95$. We note that in the thermodynamic limit $p_c$ seems to be equal to unity (see Fig.~\ref{fig::pc_ex_23}).
	\label{fig::snap}
	}
\end{figure}
Note that the above procedure, in (1), only samples unoccupied bonds, while in Ref.\,\cite{Chen11,Chen11b,Bohman04} edges are chosen uniformly at random from all edges, whether occupied or not.
In adapting random graph models to the lattice, we solely sample unoccupied bonds since in classical (random) percolation the control parameter is the fraction of occupied bonds \cite{Ziff09,Ziff10,Radicchi10,Fortunato11,Grassberger11}. In contrast, for the BFW model on the random graph, by sampling all bonds, since sufficiently many new links can be accepted asymptotically, more than one stable macroscopic cluster is obtained \cite{Chen11,Chen11b}. 
The procedure is applied iteratively while the system evolves from $t=0$, corresponding to a bond occupation fraction of $p=t/(2N)=0$, to $t=2N$ where all bonds in the system are occupied, i.\,e., $p=1$.
By construction, at each stage, the size of the largest cluster can be at most $s_\text{max}\leq k$ (see Fig.\,\ref{fig::evo2}) \cite{Bohman04}.
To illustrate the evolution of this process, we monitor in Fig.\,\ref{fig::evo2} the usual order parameter for the percolation transition, defined as the fraction of sites in the largest cluster $P_\infty=s_\text{max}/N$. One observes that the transition is delayed, compared to classical percolation where $p_c=1/2$, and the growth of the order parameter at the threshold appears to be more pronounced than in the classical case (numerical evidence discussed below). In addition, Fig.\,\ref{fig::evo2} shows the behavior of the stage per site $k/N$, which grows as the system evolves and which imposes a bound on the order parameter, $P_\infty \leq k/N$. The fraction of accepted bonds $t/u$ and the value of $g(k)$ are also shown as functions of the fraction of occupied bonds $p$. These curves illustrate the mechanism of the BFW model: if the fraction of accepted bonds were to drop below $g(k)$, the stage $k$ is incremented which decreases $g(k)$ (see insets of Fig.\,\ref{fig::evo2}). Starting from full acceptance, $t/u$ decreases, approaching the asymptotic value of $1/2$.
The mechanism promotes the homogenization of cluster sizes and delays the transition. Once percolation occurs, it is more abrupt than in the case of uniformly random bond occupation.
When $k/N=P_\infty=1$, with $p<1$, the entire set of unoccupied bonds consists of redundant bonds, that is unoccupied bonds internal to existing clusters, which can be occupied without affecting the size of clusters.
In Fig.~\ref{fig::OrderParameterPlot2D} we see the fraction of sites in the largest cluster $P_\infty$ as a function of the fraction of occupied bonds $p$ for different lattice sizes $L$.
Typical snapshots of the evolving system are shown in Fig.\,\ref{fig::snap}, for fractions of occupied bonds $p=0.5$, $0.9$, and $0.95$.

\begin{figure}
	\includegraphics[width=\columnwidth]{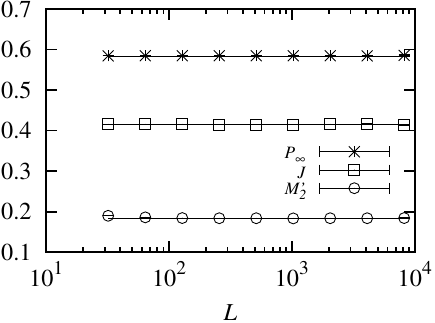}
	\caption{
	Order parameter $P_\infty$ $(\ast)$ immediately before the jump, jump size $J$ $(\square)$, and maximum of the second moment of the cluster-size distribution $M_2'$ $(\circ)$ as a function of the linear system size $L$. All three quantities are asymptotically independent of the system size.
	For large system sizes, we estimate the following values: $J=0.415\pm0.005$, $P_\infty=0.585\pm0.005$, and $M_2'=0.184\pm0.006$.
	The solid lines represent the estimated mean value for each quantity.
	These results are clear signs for a discontinuous transition \cite{Binder87,Araujo10,Schrenk11}, as previously found for the random graph \cite{Chen11,Chen11b}.
	\label{fig::jsr2d}
	}
\end{figure}

To quantify this observation, we make the following considerations. If the largest change $J$ in the order parameter $P_\infty$ is finite in the thermodynamic limit, a percolation transition is called discontinuous, as opposed to continuous transitions where the jump $J$ vanishes for $L\to\infty$ \cite{Nagler11}. Additional ways to distinguish between continuous and discontinuous transitions are to consider the behavior of the standard deviation of the order parameter and of
\begin{equation}
M_2' = M_2 - s_\text{max}^2/N \  \  ,
\end{equation}
where $M_2=\sum_i s_i^2/N$ is the second moment of the cluster-size distribution and $s_i$ is the size of cluster $i$.
Figure \ref{fig::jsr2d} shows the size independence of the largest change $\langle J \rangle_\text{j}$ and the order parameter $\langle P_\infty \rangle_\text{j}$ immediately before the biggest jump, where the average $\langle\cdot\rangle_\text{j}$ means averaging the variables when the biggest jump occurs for each realization, and the maximum of $M_2'$.
One observes that all three quantities are within error bars asymptotically independent on the system size: $J=0.415\pm0.005$, $P_\infty=0.585\pm0.005$, and $M_2'=0.184\pm0.006$. These results suggest that the transition is discontinuous \cite{Binder87,Araujo10,Schrenk11}, as was previously reported for the random graph \cite{Chen11,Chen11b}. We remark that the numerical values of $J$, $P_\infty$, and $M_2'$ are within error bars identical to the ones found for a different discontinuous percolation model in Refs.~\cite{Araujo10,Schrenk11}.

Unless indicated otherwise, on the square lattice, results have been averaged over $10^8$ samples for the smallest system size $(32^2)$ and $70$ samples for the largest one $(8192^2)$ and error bars are smaller than the symbol size. Random numbers have been generated with the algorithm proposed in Ref.\,\cite{Ziff97}. To keep track of the cluster properties as the fraction of occupied bonds increases, we have considered the labeling scheme proposed by Newman and Ziff \cite{Newman00,Newman01}, related to the Hoshen--Kopelman algorithm \cite{Hoshen76}.

\begin{figure}
	\includegraphics[width=\columnwidth]{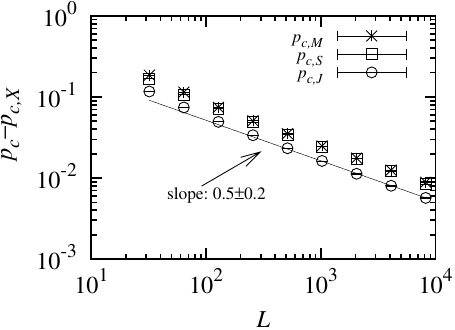}
	\caption{
	$p_c-p_{c,X}$, with $p_c=1.000\pm0.002$, as a function of the linear system size $L$ for the three estimators, the position of the jump $p_{c,J}$ $(\circ)$, the position $p_{c,M}$ $(\ast)$ of the maximum of $M_2'$, and the point $p_{c,S}$ $(\square)$ at which spanning in one specific direction occurs. Asymptotically, we observe $(p_c-p_{c,X})\sim L^{-a}$ with $a=0.5\pm0.2$.
	\label{fig::pc2d}
	}
\end{figure}

\begin{figure}
	\includegraphics[width=\columnwidth]{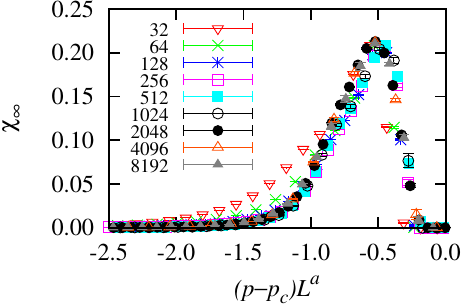}
	\caption{(Color online)
	Scaling plot for the standard deviation of the order parameter $\chi_\infty$ plotted as a function of the scaling variable $(p-p_c)L^a$, with $p_c=1$ and $a=0.5$, for different linear system sizes $L$ [$32$ $(\triangledown)$, $64$ $(\times)$, $128$ $(\ast)$, $256$ $(\square)$, $512$ $(\blacksquare)$, $1024$ $(\circ)$, $2048$ $(\bullet)$, $4096$ $(\triangle)$, and $8192$ $(\blacktriangle)$]. Note that this value of the exponent $a$ agrees within error bars with the one obtained from the asymptotic behavior of $(p_c-p_{c,X})\sim L^{-a}$, $a=0.5\pm0.2$, see Fig.\,\ref{fig::pc2d}. One observes that $\chi_\infty$ is non-zero in the thermodynamic limit, as expected for a discontinuous transition.
	\label{fig::chi2d}
	}
\end{figure}

To determine the bond occupation fraction $p_c$ at which the percolation transition occurs, we measure for every sample the position $p_{c,J}$ of the largest jump $J$, the position $p_{c,M}$ of the maximum of $M_2'$, and the point $p_{c,S}$ at which spanning in one specific direction occurs. Extrapolating these estimators to the thermodynamic limit, we find $p_c=1.000\pm0.002$. In Fig.\,\ref{fig::pc2d} we see $p_c-p_{c,X}$ as a function of the linear system size $L$, for all three estimators. Asymptotically, the points follow a power law $(p_c-p_{c,X})\sim L^{-a}$ with $a=0.5\pm0.2$. In addition, with these results we can obtain a scaling plot for the standard deviation of the order parameter $\chi_\infty$,
\begin{equation}\label{eqn::std}
\chi_\infty = \sqrt{\langle P_\infty^2 \rangle - \langle P_\infty \rangle^2} \  \  ,
\end{equation}
where $\langle\cdot\rangle$ is an average over independent realizations.
A data collapse is obtained if $\chi_\infty(p,L)$ is plotted as a function of $(p-p_c)L^a$, see Fig.\,\ref{fig::chi2d}. The fact that the maximum of $\chi_\infty$ is independent on the system size is a further evidence for a discontinuous transition \cite{Binder87,Araujo10,Schrenk11}.
Here, $\chi_\infty$ is defined as an intensive variable [see also Eq.~(\ref{eqn::std})], as the standard deviation of the largest cluster size per lattice site. In the thermodynamic limit, this quantity is zero for continuous transitions and non-zero for discontinuous transitions. For the considered system sizes, the peak height is independent of $L$. Results have been averaged over at least $1.5\times 10^3$ samples.
%

\begin{figure}
	\includegraphics[width=\columnwidth]{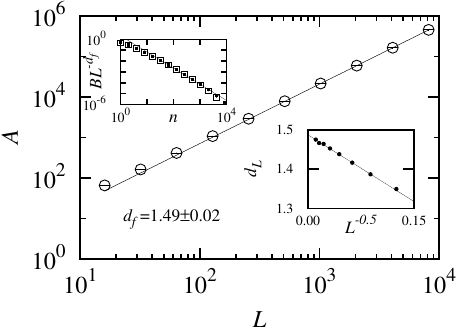}
	\caption{
	Size dependence of the number of interface bonds $A$ $(\circ)$. For large systems the interface size scales with the system size, $A\sim L^{d_f}.$ The straight line is a guide to the eye given by $0.824L^{1.47}$. In the lower right inset we see the local slopes $d_L=\log(A_L/A_{L/2})/\log(2)$ $(\bullet)$ as function of the inverse system size $L^{-0.5}$. Extrapolating to $L\to\infty$ yields an exponent of $1.49\pm0.02$. Box counting results are shown in the upper left inset: we see the rescaled number of occupied boxes $BL^{-d_f}$ as a function of the linear box size $n$ for two different system sizes: $L=8192$ ($\blacksquare$) and $L=4096$ ($\square$), the exponent is $1.49\pm0.08$. Results have been averaged over $10^8$ samples for the smallest system size $(16^2)$ and $10^3$ samples for the largest one $(8192^2)$.
	\label{fig::bfw_2_df}
	}
\end{figure}

In Fig.\,\ref{fig::snap} one observes that slightly below the percolation threshold the clusters of the BFW model are compact, but with a corrugated surface. This behavior has also been reported for another discontinuous percolation model, where the fractal dimension of the cluster perimeter was found to be consistent with the one of the bridge line in bridge percolation and the watershed \cite{Cieplak94,Cieplak96,Andrade09,Oliveira11,Andrade11,Araujo11b,Fehr09,Fehr11,Fehr11b,Araujo10,Schrenk11,Herrmann11}. Compact clusters with fractal surface have also been found in irreversible aggregation at high concentration \cite{Kolb87}. For the BFW model, we measure at the stage where the number of clusters in the system equals two, for every sample the number $A$ of bonds that would connect these two clusters if they would be occupied. The interface size \cite{Voss84} is observed to scale asymptotically with the system size, $A\sim L^{d_f}$ (see Fig.\,\ref{fig::bfw_2_df}). To estimate the fractal dimension more accurately, we plot the local slopes $d_L=\log(A_L/A_{L/2})/\log(2)$ as a function of the inverse system size $L^{-0.5}$, see lower right inset of Fig.\,\ref{fig::bfw_2_df}, which yields an exponent of $1.49\pm0.02$. This result is consistent with the fractal dimension of the interface as obtained from box counting (see upper left inset of Fig.\,\ref{fig::bfw_2_df}) which yields an exponent of $1.49\pm 0.08$. Interestingly, this value is different from the one of Ref.~\cite{Araujo10}, where $d_f=1.23\pm0.03$, but similar to the surface fractal dimension in irreversible aggregation at concentration unity, where $d_f=1.60\pm0.08$ \cite{Kolb87}.

We note that for the BFW model clusters are compact and only the perimeter is fractal, with dimension $d_f$. For $2D$ random percolation both the cluster mass and perimeter are fractals with dimension $91/48 = 1.89583\dots$ \cite{Stauffer94} and $d_f=7/4 = 1.75$ \cite{Voss84,Sapoval84}, respectively.

We also simulated the BFW model on the $2D$ triangular lattice and measured, in analogy to Fig.~\ref{fig::jsr2d}, the size of the jump $J$, the value of the order parameter $P_\infty$ immediately before the jump, and the maximum of $M_2'$ (not shown). Like on the square lattice, we found the three quantities to be asymptotically independent on the lattice size: $J=0.415\pm0.005$, $P_\infty=0.585\pm0.005$, and $M_2'=0.183\pm0.007$, within error bars identical to the results for the square lattice. This is an evidence that the BFW percolation transition is discontinuous also on the triangular lattice.
\section{\label{sec::3d}Behavior in three dimensions and multiple jumps}
%
\begin{figure}
	\includegraphics[width=\columnwidth]{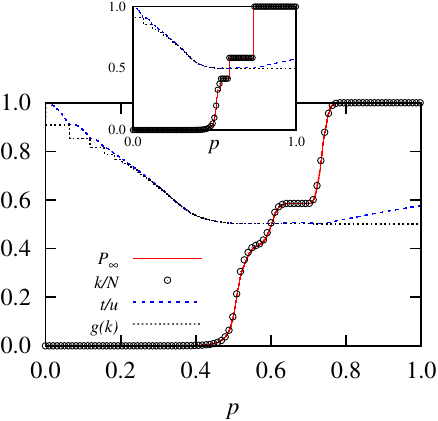}
	\caption{(Color online)
	Fraction of sites in the largest cluster $P_\infty$ (solid red line), stage per site $k/N$ $(\circ)$, occupied bonds per step $t/u$ (dashed upper blue line), and $g(k)$ (dotted lower black line) as a function of the fraction of occupied bonds $p$, for the BFW model on the simple-cubic lattice. The system size is $N=64^3$, the measured values are averages over $10^3$ samples. In the inset we see the same variables, measured for a single realization.
	\label{fig::evo3}
	}
\end{figure}
%
\begin{figure}
	\includegraphics[width=\columnwidth]{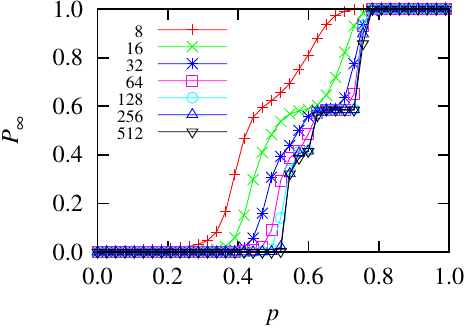}
	\caption{
	(Color online)
	Fraction of sites in the largest cluster $P_\infty$ as a function of the fraction of occupied bonds $p$ on the simple-cubic lattice for different lattice sizes $L$ (from left to right: $8$, $16$, $32$, $64$, $128$, $256$, and $512$).
	\label{fig::OrderParameterPlot3D}
	}
\end{figure}
%
\begin{figure}
	\begin{tabular}{l}
	(a)\\
	\includegraphics[width=\columnwidth]{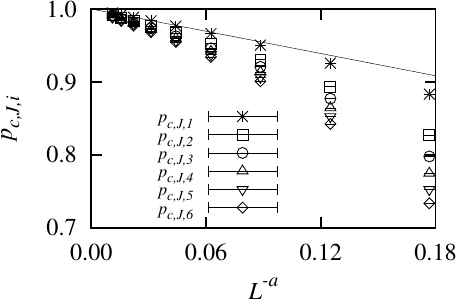}\\
	(b)\\
	\includegraphics[width=\columnwidth]{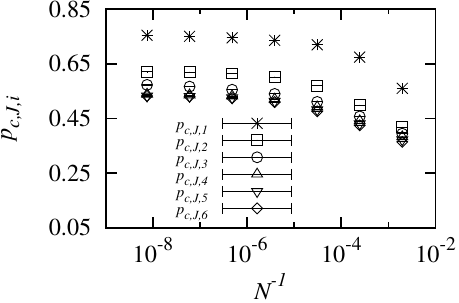}\\
	\end{tabular}
	\caption{
	Bond occupation fractions $p_{c,J,i}$ at which the six largest jumps in the order parameter occur for the BFW model as a function of the inverse system size, for (a) the square lattice and (b) the simple-cubic lattice. While for two dimensions all $p_{c,J,i}$ can be extrapolated to the same value $p_c=1.000\pm0.002$, in three dimensions four distinct jumps are visible within error bars for $L\to\infty$ (compare also Fig.\,\ref{fig::evo3}). For the square lattice, $a=0.5\pm0.2$ and the solid line is a guide to the eye. On the simple-cubic lattice we find $p_{c,1}=0.756\pm0.005$, $p_{c,2}=0.623\pm0.007$, $p_{c,3}=0.578\pm0.006$, and $p_{c,4}=0.539\pm0.008$. For (b) results have been averaged over $10^8$ samples for the smallest system size $(8^3)$ and $40$ samples for the largest one $(512^3)$.
	\label{fig::pc_ex_23}
	}
\end{figure}
In Sec.\,\ref{sec::2d} we considered the BFW model on the square lattice and found evidence for a discontinuous transition.
We proceed to analyze the BFW model on the simple-cubic lattice. In Fig.\,\ref{fig::evo3} we see the analog of Fig.\,\ref{fig::evo2} for the simple-cubic lattice, namely, the fraction of sites in the largest cluster $P_\infty$, stage per site $k/N$, occupied bonds per step $t/u$, and $g(k)$ as a function of the fraction of occupied bonds $p$. One observes that the fraction of sites in the largest cluster $P_\infty$ grows stepwise (see also Fig.~\ref{fig::OrderParameterPlot3D}), which hints at the presence of multiple stable components, as reported for the random graph \cite{Chen11,Chen11b}. As on the square lattice (see Fig.\,\ref{fig::evo2}), the size of the largest cluster is bounded by the stage $k$. The fraction of accepted bonds $t/u$ decreases initially, ultimately reaching the asymptotic value of $g(k)=1/2$ before increasing slightly once only one cluster remains. In stages where $s_\text{max}$ is essentially constant, the fraction of accepted bonds increases due to the addition of redundant bonds.

\begin{figure}
	\includegraphics[width=\columnwidth]{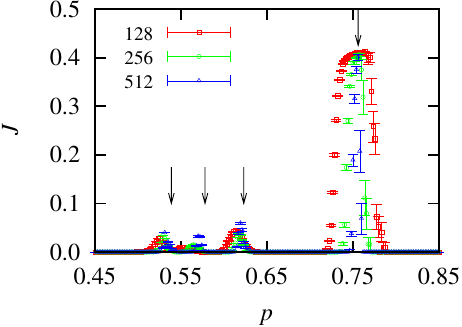}
	\caption{(Color online)
	Maximum change $J$ in the order parameter in the BFW model on the simple-cubic lattice as a function of the fraction of occupied bonds $p$ for different linear system sizes $L$ [$128$ $(\square)$, $256$ $(\circ)$, and $512$ $(\triangle)$: peaks become narrower with increasing $L$]. The arrows indicate the positions of the four largest jumps, extrapolated to $L\to\infty$: $p=0.756$, $0.623$, $0.578$, and $0.539$ (see Fig.~\ref{fig::pc_ex_23}). The range between zero and unity for $p$ was divided into $512$ bins of equal size and in every bin the maximum of the occurred changes in the order parameter was measured. Results have been averaged over $1.6\times10^5$ samples for the smallest system size $(128^3)$ and $2.4\times10^3$ samples for the largest one $(512^3)$.
	\label{fig::NonZeroSmaxChange3DFull}
	}
\end{figure}

To analyze this effect in detail, we proceed as follows. For every sample we measure the $N_J$ largest increases in the order parameter $P_\infty$, denoted as $J_i$, and the fraction of occupied bonds $p_{c,J,i}$ at which they occur, and average over many samples. This definition contains the above ones of $J$ and $p_{c,J}$, since $J=J_1$ and $p_{c,J}=p_{c,J,1}$. Let us consider the six largest increases in the order parameter, i.\,e., $N_J=6$.
In the $2D$ square lattice, only one jump occurs: one observes that, extrapolated to $L\to\infty$, the occupation fractions $p_{c,J,i}$ are identical within their error bars, see Fig.\,\ref{fig::pc_ex_23}(a), and the corresponding jump sizes $J_i$ are macroscopic (not shown). This means that in the thermodynamic limit the fraction of bonds that is added between two macroscopic increases of the order parameter is zero. In this sense, for two dimensions, all jumps happen at the same instant $p=p_c$. In contrast, as shown in Fig.\,\ref{fig::pc_ex_23}(b), in three dimensions, the macroscopic jumps occur at different bond occupation fractions $p_{c,J,i}$. This is consistent with multiple stable clusters emerging and coexisting when added bonds are mainly redundant ones.
In Fig.~\ref{fig::NonZeroSmaxChange3DFull} we see the jump size $J$ as a function of the fraction of occupied bonds $p$ for the simple-cubic lattice.
We divide the considered range of bond occupation fractions $p$ into $512$ bins of equal size. In each bin the maximum of the changes in the order parameter is measured and averaged over several samples. In the plot, one sees four peaks with heights increasing or being constant with increasing lattice size. Since these peaks are well-separated and do not seem to approach each other for increasing lattice sizes, this corresponds to four macroscopic jumps occurring at distinct values of $p$. This is consistent with the results shown in Fig.~\ref{fig::pc_ex_23}(b) and \ref{fig::evo3}: the arrows in Fig.~\ref{fig::NonZeroSmaxChange3DFull} indicate the positions of the four largest increases in the largest cluster size, as obtained in Fig.~\ref{fig::pc_ex_23}(b).
We remark that the number of jumps measured at distinct fractions of occupied bonds depends on the resolution, which in Fig.~\ref{fig::NonZeroSmaxChange3DFull} is related to the bin size. An interesting possibility would be that the fraction of sites in the largest cluster $P_\infty$ as function of the fraction of occupied bonds $p$ actually behaves like a Devil's Staircase, with an infinite hierarchy of jumps.
In the next section, we study the tree-like version of the model where solely merging bonds are occupied, i.\,e., bonds connecting edges belonging to different clusters.
\section{\label{sec::llbfw}Tree-like BFW model}
%
\begin{figure}
	\begin{tabular}{l}
	(a)\\
	\includegraphics[width=.92\columnwidth]{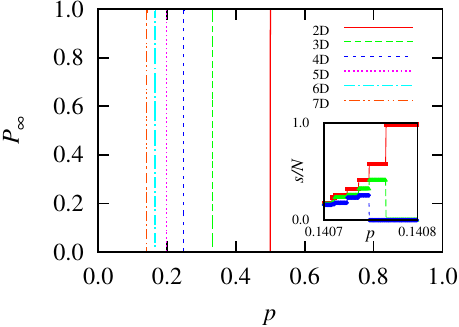}\\
	(b)\\
	\includegraphics[width=.92\columnwidth]{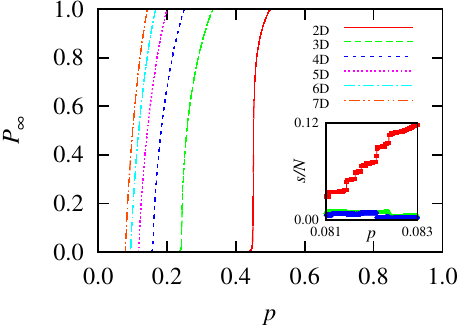}\\
	\end{tabular}
	\caption{
	(Color online)
	Fraction of sites in the largest cluster $P_\infty$ as a function of the fraction of occupied bonds $p$ for (a) the tree-like BFW model and (b) tree-like classical percolation on hypercubic lattices of dimension two to seven (from right to left with increasing dimension). For the BFW model the transition is abrupt. In the classical case, the transition is continuous and the thresholds are compatible with results for the number of clusters at the critical point \cite{Ziff97b,Temperley71,Baxter78}. The insets show the fraction of sites $s/N$ in the three largest clusters [1st ($\square$, upper curve), 2nd ($\triangle$, middle curve), and 3rd ($\circ$, lower curve)] as a function of the fraction of occupied bonds $p$ for single typical realizations of (a) the tree-like BFW model and (b) tree-like classical percolation on the hypercubic lattice of dimension seven of size $6^7$. While for the BFW model the largest cluster grows by overtaking, i.\,e., the second largest cluster overtakes the largest one which becomes the second largest one, for classical percolation the largest cluster grows directly by merging with smaller clusters. For the main plots, the system sizes $L^d$ are $2048^2$, $161^3$, $45^4$, $21^5$, $13^6$, and $9^7$ for both models. Results have been averaged over at least $2\times10^3$ samples.
	\label{fig::op27}
	}
\end{figure}
The BFW model as defined in Sec.\,\ref{sec::2d} seems to exhibit a discontinuous transition on the square and simple-cubic lattices. We found that in three dimensions multiple jumps of the order parameter occur at distinct bond occupation fractions in the thermodynamic limit, which can only occur when redundant bonds are occupied. We now consider a modified version of the BFW model where redundant bonds are never occupied, i.\,e., bonds connecting sites that are part of the same cluster are not sampled and, therefore, all clusters are tree-like. This version was also considered for random graphs in Ref.\,\cite{Chen11} under the name {\it restricted BFW model}.

Figure \ref{fig::op27} shows $P_\infty$ as a function of the fraction of occupied bonds $p$ for the tree-like BFW model and, for the sake of comparison, we also include the results for tree-like classical percolation. We observe that under this constraint, for the BFW model the transition seems abrupt in all considered dimensions, whereas in the classical case the transition is continuous and the percolation thresholds are consistent with the values expected for the number of clusters at the critical point (see Ref.\,\cite{Ziff97b,Temperley71,Baxter78} and references therein). In particular, at $p_c$, the number of clusters per site $(N_c/N)$ in tree-like percolation is $N_c/N=1-p_cd$, which on the square lattice, yields $p_c\approx(7-3\sqrt{3})/4$ \cite{Ziff97b,Temperley71,Baxter78}. The insets in Fig.\,\ref{fig::op27} show the fraction of sites $s/N$ in the three largest clusters as a function of the fraction of occupied bonds $p$ for single realizations of the tree-like BFW model [Fig.\,\ref{fig::op27}(a)] and tree-like classical percolation [Fig.\,\ref{fig::op27}(b)]. One observes that for the BFW model the size of the largest cluster grows due to two smaller clusters merging together to become the new largest (growth by overtaking), while in the classical case the largest component mainly grows by adding small pieces \cite{Nagler11}. In the inset of Fig.\,\ref{fig::op27}(a) the symbols corresponding to the largest cluster form the curve on top. When a jump occurs, the size of the largest cluster changes by overtaking: for example at the second last jump shown in the inset, the second largest cluster merges with the third largest cluster and becomes the largest cluster, while the previous largest cluster is then the second largest one. For the complete graph it has been shown that all significant growth of the largest cluster size is due to overtaking \cite{Chen11b}.

\begin{figure}
	\includegraphics[width=\columnwidth]{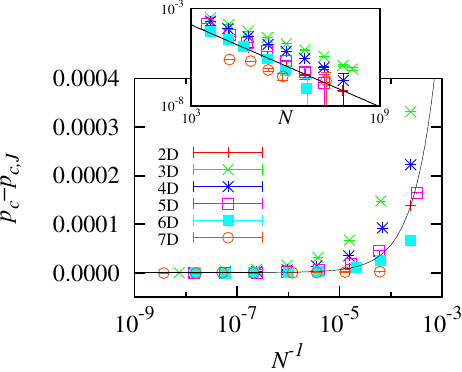}
	\caption{
	(Color online)
	Percolation threshold estimator $p_c-p_{c,J}$ as a function of the inverse system size $N^{-1}$ for the tree-like BFW model on the hypercubic lattice of dimensions two to seven. Extrapolating to $L\to\infty$, we find $p_c=0.500\pm0.001$, $0.333\pm0.002$, $0.248\pm0.002$, $0.198\pm0.003$, $0.165\pm0.004$, and $0.141\pm0.009$. Note that $p_c$ seems to be $1/d$.
	In the inset we see the same data for $p_c-p_{c,J}$, plotted as a function of $N=L^d$ in double-logarithmic scale, where the solid line is a guide to the eye of the form $0.018N^{-0.7}$.
	Results have been averaged over at least $7.8\times 10^6$ samples for the smallest system size ($64^2$, $16^3$, $8^4$, $5^5$, $4^6$, and $4^7$) and over at least $100$ samples for the largest one ($8192^2$, $512^3$, $91^4$, $37^5$, $20^6$, and $16^7$).
	\label{fig::pc27}
	}
\end{figure}
\begin{figure}
	\includegraphics[width=\columnwidth]{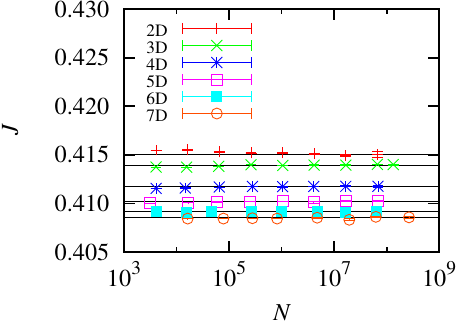}
	\caption{
	(Color online)
	The largest change in the order parameter: $J$ is plotted as a function of $N=L^d$ for the tree-like BFW model on hypercubic lattices of dimension two to seven. Asymptotically, the jump is independent on system size, indicating a strongly discontinuous transition. The solid lines are guides to the eye indicating the following constant values: $0.415\pm0.005$, $0.414\pm0.002$, $0.412\pm0.003$, $0.410\pm0.003$, $0.409\pm0.002$, and $0.409\pm0.002$, for dimension two to seven.
	\label{fig::jump27}
	}
\end{figure}

We now analyze the behavior of the tree-like BFW model on hypercubic lattices. To estimate the percolation thresholds for the infinite system, we consider the estimator $p_J$, i.\,e., the bond occupation fraction at which the largest change in the order parameter $P_\infty$ occurs (see Sec.\,\ref{sec::2d}) and study its size dependence. For dimensions two to seven we obtain, respectively, $p_c=0.500\pm0.001$, $0.333\pm0.002$, $0.248\pm0.002$, $0.198\pm0.003$, $0.165\pm0.004$, and $0.141\pm0.009$ (see Fig.\,\ref{fig::pc27}). These values are within error bars equal to $1/d$. In Ref.\,\cite{Schrenk11} it was shown that for any percolation model on hypercubic lattices of dimension $d$ with finite number of clusters at the threshold $p_c \geq 1/d$. If, in addition, the models are tree-like, $p_c=1/d$ was shown. The values of the jump $J$ are summarized in Fig.\,\ref{fig::jump27}, where $J$ is plotted as a function of the system size $N$ for the tree-like BFW model in all considered dimensions. One observes that the jump value converges towards a constant value for large system sizes, which is a strong evidence of a discontinuous transition \cite{Nagler11}.

\begin{figure}
	\includegraphics[width=\columnwidth]{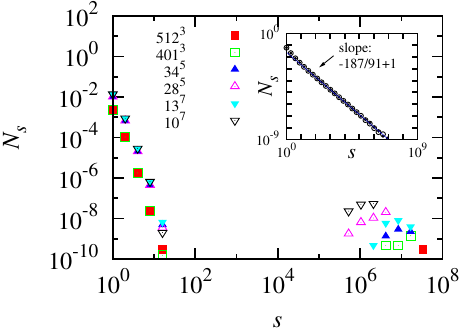}
	\caption{
	(Color online)
	Accumulated cluster-size distribution $N_s$ as a function of the cluster size $s$ for the tree-like BFW model on hypercubic lattices of dimension three $(\blacksquare, \square)$, five $(\blacktriangle, \triangle)$, and seven $(\blacktriangledown, \triangledown)$. $N_s=\sum_{r=s}^{2s-1}n_r$, where $n_s$ is the number of clusters of size $s$ per lattice site $N=L^d$, excluding the largest cluster of size $s_\text{max}$. One observes that the distribution falls off sharply for small cluster sizes and has a region on the right, qualitatively similar to the behavior on the random graph \cite{Chen11,Chen11b}. For comparison, the inset shows $N_s$ as a function of $s$ for tree-like classical percolation on the square lattice at $p_c=(7-3\sqrt{3})/4$ \cite{Ziff97b,Temperley71,Baxter78}. Results are shown for system sizes of $8192^2$ $(\circ)$ and $16384^2$ $(\bullet)$. Asymptotically, the points follow a power law, $N_s\sim s^{1-\tau}$, where $\tau=187/91$. All results have been averaged over $100$ samples.
	\label{fig::csd}
	}
\end{figure}

In the following we investigate the cluster-size distribution at the threshold for high dimensions.
In Fig.\,\ref{fig::csd} we see a plot of the number of clusters of size $s$ per site $n_s$ as a function of the cluster size $s$ for hypercubic lattices of dimension three, five, and seven.
To reduce statistical fluctuations, we considered the accumulated number of clusters $N_s$, defined as $N_s=\sum_{r=s}^{2s-1}n_r$.
It can be observed that the cluster-size distribution is composed of two parts as the dimension increases, in agreement with the behavior on the random graph \cite{Chen11,Chen11b}. The number of small clusters at the threshold increases with dimension. Thus the cluster-size distribution for the tree-like BFW model behaves in a different way than in classical percolation at the critical point, where asymptotically a power law holds, $n_s\sim s^{-\tau}$ or $N_s\sim s^{1-\tau}$ (see inset of Fig.\,\ref{fig::csd}).

Besides finding some evidence that the tree-like BFW model on the hypercubic lattice exhibits a discontinuous transition, the results presented in this section also show that plateaus in the evolution of the fraction of sites in the largest cluster $P_\infty$ on the simple-cubic lattice (see Fig.\,\ref{fig::evo3}) are due to redundant bonds, which are not present in the tree-like case, in agreement with what has been reported for the random graph \cite{Chen11,Chen11b}. However, on the two-dimensional square lattice, the jumps occur at one unique fraction of occupied bonds, irrespective of whether the tree-like version is considered or not.
This result is in agreement with the results of Moreira {\it et al.}, showing that to obtain a discontinuous percolation transition in two dimensions, the control of internal/redundant bonds is irrelevant \cite{Moreira10}.
\section{\label{sec::end}Final Remarks}
We analyzed the BFW model \cite{Bohman04,Chen11,Chen11b} on the lattice and found numerical evidence for a strongly discontinuous transition \cite{Nagler11} in two and three dimensions. The surface fractal dimension of the BFW model in two dimensions was measured to be $d_f=1.49\pm0.02$. Interestingly, this is different from the fractal dimension reported for another discontinuous percolation model, $1.23\pm0.03$ \cite{Araujo10,Herrmann11}, but similar to the one found in irreversible aggregation at high concentration, $1.60\pm0.08$ \cite{Kolb87}. The BFW model on the simple-cubic lattice was found to exhibit multiple distinct jumps of the largest cluster size in the thermodynamic limit. We further analyzed the tree-like BFW model on hypercubic lattices of dimension two to seven, finding that the jump size is asymptotically independent on system size and determined the percolation thresholds. These results imply that the coexistence of multiple stable components in the BFW model is due to redundant bonds.
This can be seen from the fact that the tree-like version of the BFW model only exhibits one jump while multiple jumps at distinct fractions of occupied bonds appear only for the simple-cubic lattice when redundant bonds are also sampled.
We also considered the evolution of the cluster-size distribution at the threshold with increasing dimension.

The results presented here are consistent with the behavior of the BFW model on the random graph \cite{Chen11,Chen11b}. The mechanisms leading to discontinuous percolation in this case are compatible with the required ones according to Ref.\,\cite{Moreira10}, namely, the homogenization of cluster sizes and the control of merging bonds. Identifying percolation models with discontinuous transitions is interesting since the original proposal \cite{Achlioptas09} and related models have been shown, both analytically and numerically, to yield continuous transitions in the thermodynamic limit \cite{Riordan11}. Nevertheless, these initial models are characterized by a macroscopic jump for finite networks even with $10^{18}$ nodes \cite{daCosta10,Nagler11}. Note that, the Internet and the WWW have less than $10^{12}$ nodes.

It would be interesting to understand the finite-size scaling of discontinuous percolation models in a more rigorous analytical manner since, in the presence of redundant bonds, it differs from the one of equilibrium first-order phase transitions \cite{Binder87}. In addition, interesting fractal dimensions emerged in the investigation of cluster perimeters in discontinuous percolation \cite{Cieplak94,Cieplak96,Andrade09,Oliveira11,Andrade11,Araujo11b,Fehr09,Fehr11,Fehr11b,Schrenk11} which could also be studied for the BFW model in dimensions higher than two.
\begin{acknowledgments}
We acknowledge financial support from the ETH Risk Center.
We also acknowledge financial support from the Brazilian agencies CNPq, CAPES and FUNCAP, the Pronex grant CNPq/FUNCAP, the Defense Threat Reduction Agency award HDTRA1-10-1-0088, and the Army Research Laboratory award W911NF-09-2-0053.
\end{acknowledgments}

\bibliography{bfw}

\end{document}